\documentclass[10pt,aps,preprintnumbers,superscriptaddress,nofootinbib,aps,prd,floatfix,twocolumn]{revtex4-2}
\pdfoutput=1 
\usepackage{graphicx}%
\usepackage{amsthm} 
\usepackage{amsmath,amssymb,physics}
\usepackage{hyperref}
\usepackage{url} 
\usepackage{lineno} %
\usepackage{xspace}  %
\usepackage{braket}  %
\usepackage{wrapfig}  %
\usepackage{xcolor} 
\usepackage{comment} %
\usepackage{array} %
\usepackage{%
            soul}
\usepackage{longtable}
\usepackage{xparse}
\usepackage{subfigure}

 \usepackage{setspace}

\newcommand{\Fantomas}{\texttt{Fant\^omas}\xspace}  %
\newcommand{\meta}{metamorph\xspace}
\newcommand{\Bezier}{B\'ezier\xspace}
\newcommand{\METAPDF}{\texttt{METAPDF}\xspace}
\newcommand{\xFitter}{\texttt{xFitter}\xspace}  %
\newcommand{\FantoPDF}{\texttt{Fanto10\_15}\xspace}  %
\newcommand{\nFantoPDF}{\texttt{Fanto10\_n15}\xspace}  %
\newcommand{\nCTEQ}{\texttt{nCTEQ15}\xspace}

\usepackage{pythonhighlight}
\usepackage{listings}

\definecolor{codegreen}{rgb}{0,0.6,0}
\definecolor{codegray}{rgb}{0.5,0.5,0.5}
\definecolor{codepurple}{rgb}{0.58,0,0.82}
\definecolor{backcolour}{rgb}{0.95,0.95,0.92}

\lstdefinestyle{mystyle}{
    backgroundcolor=\color{backcolour},   
    commentstyle=\color{codegreen},
    keywordstyle=\color{magenta},
    numberstyle=\tiny\color{codegray},
    stringstyle=\color{codepurple},
    basicstyle=\ttfamily\footnotesize,
    breakatwhitespace=false,         
    breaklines=true,                 
    captionpos=b,                    
    keepspaces=true,                 
    numbers=left,                    
    numbersep=5pt,                  
    showspaces=false,                
    showstringspaces=false,
    showtabs=false,                  
    tabsize=2,
    xleftmargin=0.5in,
    xrightmargin=0.5in
}

\NewExpandableDocumentCommand\mcc{O{1}m}
    {\multicolumn{#1}{l}{#2}}
\usepackage{multirow}
\newcolumntype{C}[1]{>{\centering\let\newline\\\arraybackslash\hspace{0pt}}m{#1}}

\definecolor{nicegreen}{rgb}{0.,0.5,0.}

\newcommand{\orcid}[1]{\,\href{https://orcid.org/#1}{\includegraphics[width=9pt]{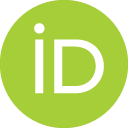}}}

\newcommand{\orcidPN}{0000-0003-3732-0860} %
\newcommand{\orcidAC}{0000-0001-8906-2440} %
\newcommand{\orcidMAX}{0009-0003-0139-4072} %
\newcommand{\orcidLK}{0009-0007-5639-0350} %

\def\msu{Department of Physics and Astronomy,
Michigan State University, East Lansing, Michigan 48824, USA}
\def\smu{Department of Physics, Southern Methodist University, Dallas, TX 75275-0175, USA}
    
\def\unam{\mbox{Instituto de F\'isica,
  Universidad Nacional Aut\'onoma de M\'exico, Apartado Postal 20-364,
  01000 Ciudad de M\'exico, Mexico}}

 \modulolinenumbers[5]

\begin{document}
\author{Lucas Kotz\orcid{\orcidLK}}\affiliation{\smu}
\author{Aurore Courtoy\orcid{\orcidAC}}
\email{aurore@fisica.unam.mx}\affiliation{\unam}
\author{Pavel Nadolsky\orcid{\orcidPN}}\email{nadolsky@smu.edu}\affiliation{\msu}
\author{Maximiliano Ponce-Chavez\orcid{\orcidMAX}}\affiliation{\msu}

\renewcommand*{\thefootnote}{\arabic{footnote}}

\begin{abstract}
 We present \texttt{Fanto10}, a new ensemble of NLO error parton distribution functions (PDFs) in a charged pion that provides the most detailed estimate of uncertainties from experimental, theoretical, and methodological sources in order to enable faithful comparisons against upcoming precision experiments and {\it ab initio} QCD predictions. For the first time, the \texttt{Fanto10} PDFs quantify two important types of uncertainties, arising from in-depth exploration of feasible functional forms of pion PDFs and from the nuclear PDFs describing the internal composition of the tungsten target in the key E615 data set for pion-nuclear Drell-Yan process. Accounting for these uncertainties modifies physics conclusions 
of more restrictive analyses about the pion's gluon and quark sea composition. These advancements are made by employing the recently developed C++ framework \Fantomas for systematic exploration of epistemic sources of PDF uncertainties, such as those arising from the choice of the PDF functional forms and sampling over methodological choices and third-party inputs, e.g., the nuclear PDF uncertainty in the presented case. The framework employs polynomial universal approximators (B\'ezier curves) to parametrize diverse PDF functional forms and reduce biases associated with the PDF parametrization choice. Its other component combines PDF ensembles from multiple preliminary fits into a single published ensemble of Hessian error PDFs. We review key steps of the \Fantomas methodology, properties of \texttt{Fanto10} PDFs, and implications for future studies.
\end{abstract}

\title{ \Fantomas: epistemic and nuclear uncertainties for the parton distributions of the pion }

\maketitle

{\it Introduction.}
The structure of pions is a subject of active inquiry thanks to the advent of new experiments that use either pion beams~\cite{Andrieux:DIS24} or virtual pion targets~\cite{AbdulKhalek:2021gbh, Arrington:2021biu, Accardi:2023chb}, a challenging task owing to the pion's instability and small mass. Although pions seemingly possess a simpler constituent quark structure compared to nucleons, their internal dynamics remains intriguing due to imprint of chiral symmetry breaking on formation of pseudo-Goldstone bosons. Key open questions include the gluon content of pions at a given resolution scale and the role of quarks and gluons in pion's mass emergence through a combination of QCD-driven and electroweak Higgs mechanisms. Insights into both questions can be gleaned from exploring the flavor composition and shapes of parton distribution functions (PDFs) of the pion in high-energy collisions. 
To interpret these results, one must understand and quantify uncertainties on pion PDFs. 

Recent analyses of pion PDFs offer a valuable testing ground for new methods in solving inverse problems, taking upon statistics issues of significant depth. 
The JAM collaboration has led many an effort~\cite{Barry:2018ort, Cao:2021aci,Barry:2021osv, JeffersonLabAngularMomentumJAM:2022aix}, accounting for data inputs from leading-neutron deeply inelastic scattering (DIS)
and even first constraints from lattice QCD, as well as threshold resummation. The \xFitter collaboration published an independent Hessian-based fit~\cite{Novikov:2020snp} of the pion PDFs that served as a progenitor for our study. 
On the other hand, progress has been made in the context of nucleon PDF fits~\cite{Courtoy:2022ocu,Kotz:2023pbu, Kriesten:2023uoi,  Costantini:2024wby} towards quantification of uncertainties coming from the model and methodology, referred to as {\it epistemic}, as in the sense of {\it systematic}, uncertainties, as opposed to the contribution coming from {\it aleatoric}, or {\it statistical} uncertainties. Recent pion analyses~\cite{Barry:2018ort, Cao:2021aci,Barry:2021osv, JeffersonLabAngularMomentumJAM:2022aix,Novikov:2020snp} also explored some aspects of quantification of systematic uncertainties, yet leaving room for further advancements.

It is with the aim of achieving PDF determination with comprehensive uncertainties that our team determined the pion PDFs $f_a(x,Q_0)$ along a new methodology, \Fantomas, introduced to systematically explore their functional dependence on the momentum fraction $x$ at the initial scale $Q_0$ of QCD evolution. Our first publication \cite{Kotz:2023pbu} documenting this method described a versatile representation for the PDF, called a \meta,  based on a polynomial approximator chosen to be a \Bezier curve. Metamorphs have direct physics interpretation and reduce computational needs for generation and refitting of multiple PDF parametrizations when estimating the associated uncertainty, which contributes a large part, sometimes a half or more \cite{Hou:2019efy}, of the total PDF uncertainty. Traditional polynomial forms of fitted PDFs often can be converted into the standard format of metamorphs, then multiple fits based on descendant functional forms can be automated and parallelized. 

 The \Fantomas approach allows one to reduce the parametrization bias, associated with arbitrariness in choosing the functional forms~\cite{Courtoy:2022ocu}, by fitting and combining various equally acceptable PDF solutions~\cite{Gao:2013bia,UsTechnical:2025}. We, the \Fantomas team, have shown that consequently the uncertainty tends to increase {\it w.r.t.} other analyses for the pion PDFs based on restrictive parametrizations, in $x$ intervals where there are little to no data.  Such behavior, also present with neural-network PDF models \cite{Ball:2010de,Ball:2014uwa}, here is obviated by an interpretable polynomial representation. As a demonstration of possible physics implications, the metamorph-based fit \cite{Kotz:2023pbu} implemented in \xFitter revealed a strong anti-correlation between the pion's sea and gluon distributions, weakening the earlier findings \cite{Barry:2018ort, Novikov:2020snp, JeffersonLabAngularMomentumJAM:2022aix} of a large gluon content in the pion, as characterized by the gluon's momentum fraction. 
 
 After the publication of Ref.~\cite{Kotz:2023pbu}, we implemented the updated \Fantomas methodology in a public C++ code that will be released together with a companion manual~\cite{UsTechnical:2025}.\footnote{Additional aspects of our pion fit, such as the exploration of constraints of individual data sets on pion PDFs obtained with the method of $L_2$ sensitivities \cite{Hobbs:2019gob,Jing:2023isu}, are documented in a PhD thesis~\cite{KotzThesis:2025}.}

In this Letter, we elaborate on the resulting ensemble of pion PDFs named $\texttt{Fanto10}$, to point out that, together with the bias-suppressed implementation of parametrizations, for the first time it incorporates an uncertainty coming from the PDF in a nuclear target, relevant for the data set of the pion fit obtained in Drell-Yan pair production in pion-tungsten scattering. Implementation of both new types of uncertainties is achieved through the \METAPDF approach for combination of PDF ensembles, also used in the PDF4LHC combinations \cite{Butterworth:2015oua, PDF4LHCWorkingGroup:2022cjn}. Originally proposed in~\cite{Gao:2013bia}, the \texttt{META} analysis transforms all PDF solutions into a Monte-Carlo (MC) replica ensemble, and then converts it into a Hessian eigenvector ensemble. Here we follow this approach to incorporate multiple unrelated uncertainties into the final error ensemble.

{\it The probability of models emerges from their sampling.}
As the epistemic probability on the space of functional forms of PDFs is unknown {a priori}, the philosophy of the \Fantomas method is to learn it incrementally through representative sampling \cite{Courtoy:2022ocu} of PDF models according to several guiding principles. First, each accepted PDF fit must describe well the fitted data, with less agreeing models disfavored in the final combination. Second, simpler models are preferred under other equal conditions to facilitate generalization for future predictions. Third, models selected for the final combination must be representative of the available ensemble of good solutions. For example, when combining models with the same goodness of fit, it makes sense to keep diverse, relatively equidistant models according to their separation in physical PDF space, as clustering of the models in one corner of parameter space is not expected from their prior probability. Fourth, models must be conditionally independent and exchangeable when included in the combination. In this case, and in line with de Finetti’s representation theorem for exchangeable sequences~\cite{definetti1974theory}, PDF ensembles entering the final combination can be viewed as sampled from some epistemic probability distribution associated with an unknown latent parameter $\theta$ assigned to each sampled model. The final PDF error band then accounts for the aleatoric uncertainty, associated with random fluctuations of data, as well as the emergent epistemic one that reflects sampling over the explored PDF models.

{\it Model combination in the \texttt{Fanto10} fit} illustrates application of these general principles according to a simplified procedure that is sufficient at the current, still low, accuracy of pion data.\footnote{Refinements in the combination procedure are planned for precise nucleon fits.} Ref.~\cite{Kotz:2023pbu} details settings of our pion fit. It is performed at the next-to-leading order (NLO) in QCD and introduces three independent parametrizations for the valence, SU(3)-symmetric quark sea, and gluon distributions, denoted by $xV(x)$, $xS(x)$, $xg(x)$ and given by a metamorph with up to four free parameters each. These PDFs are constrained by fixed-target measurements of Drell-Yan (DY) pair and direct photon production, covering $x\gtrsim 0.07$ and $Q\lesssim 15$ GeV, and by 29 points from leading-neutron DIS, covering $ 8.16\cdot 10^{-4} < x_\pi < 0.0765$. Among these, the E615 and NA10 data sets on pion-tungsten DY process dominate the large-$x$ constraints, and hence the pion PDFs are sensitive to the assumed nuclear ones, which we take from the \nCTEQ NLO ensemble \cite{Kovarik:2015cma}. 

We start by obtaining about 100 candidate fits to pion data with varied PDF parametrization forms, cf. Ref.~\cite{Kotz:2023pbu}, and for the same central ensemble of \texttt{nCTEQ15FullNuc\_184\_74} nuclear PDFs, referring to a tungsten-184 target, and \texttt{nCTEQ15FullNuc\_1\_1} for the proton baseline in WA70 direct photon production. We then rank fits, or models $M$ according to an objective function defined as $\chi^2=\chi^2_{D|M}+\chi^2_{\rm M}$, where $\chi^2_{D|M}\equiv -2 \ln\left(P(D|M)\right)$ is the log-likelihood for the fitted pion data $D$ given $M$, and $\chi^2_{M}\equiv -2 \ln\left(P(M)\right)$ represents the possible contribution from priors constraining $M$. In the context of the \Fantomas fit, 
$M$ corresponds to specific settings of the metamorphs, and $\chi^2_{M}$ may be imposed to prevent runaway free parameters in some fits given the paucity of data, {\it e.g.} due to the insufficiently constrained asymptotic power $B_{S}$ of the sea PDF $xS(x,Q_0)$, growing as $x^{B_S}$ at $x\to 0$, see Sec.~3 in \cite{Kotz:2023pbu}.

\begin{figure}
    \centering
    \includegraphics[width=0.75\linewidth]{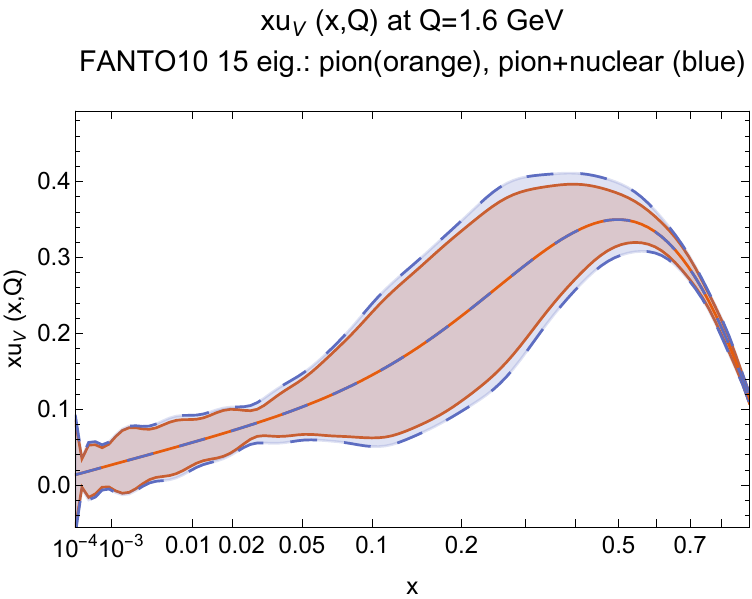}
    \includegraphics[width=0.75\linewidth]{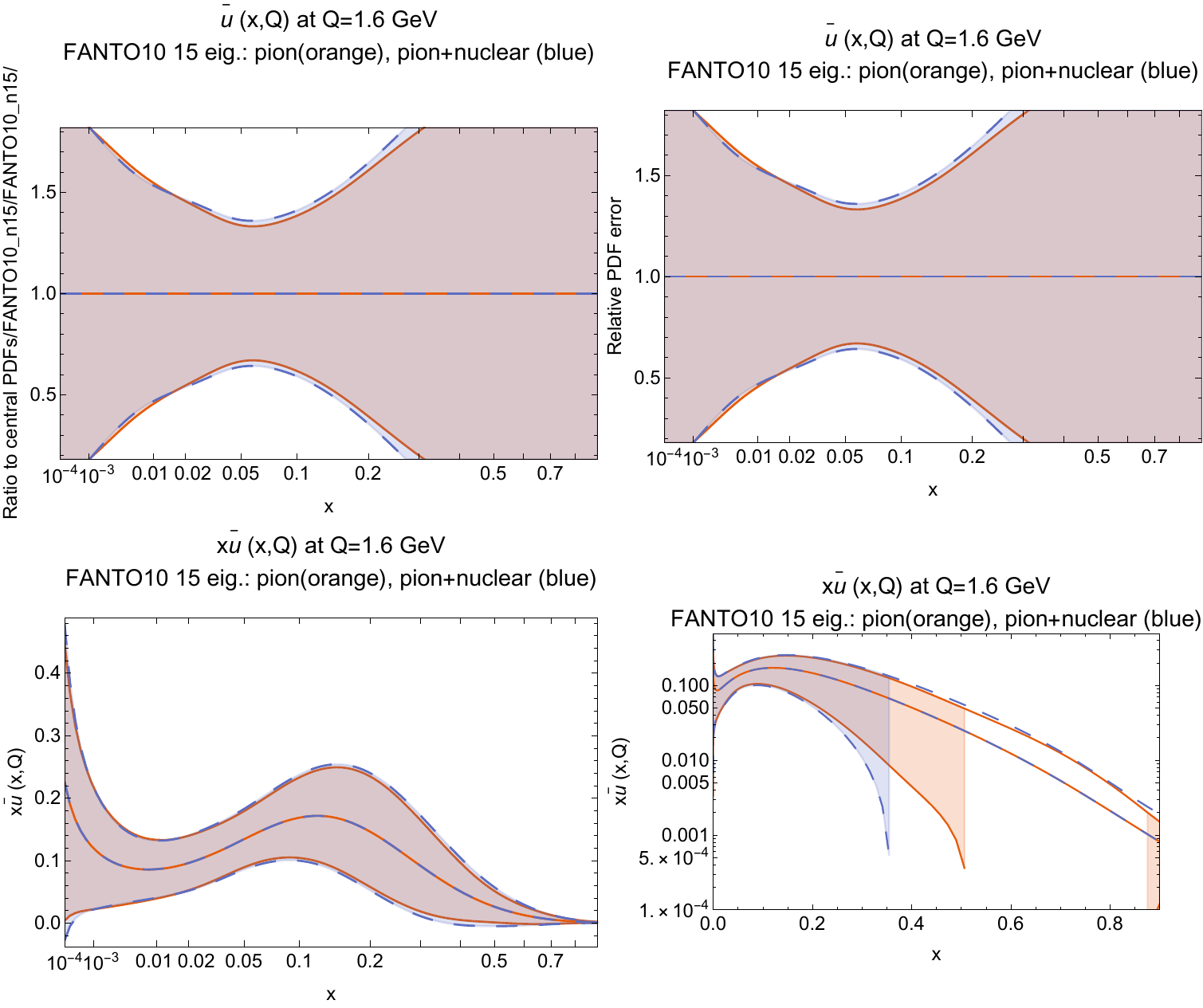}
    \includegraphics[width=0.75\linewidth]{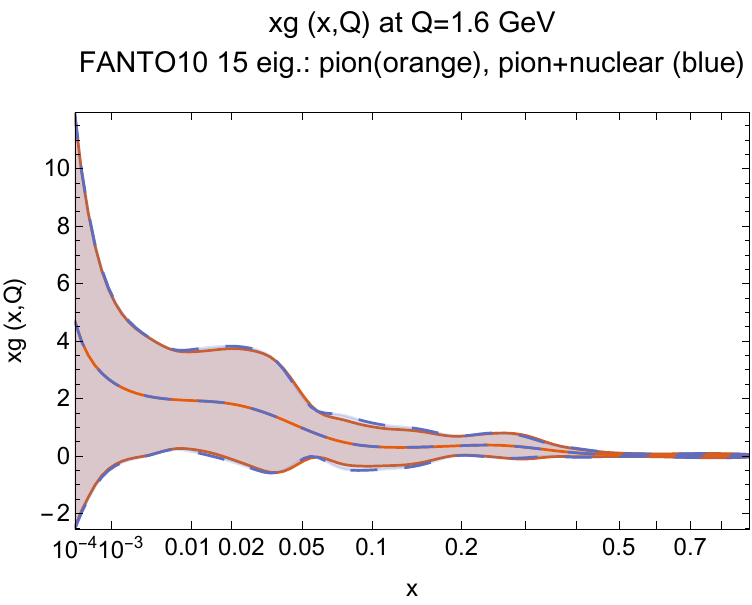}
    \caption{The aleatoric+epistemic (orange) and aleatoric+epistemic+nuclear (blue) uncertainty bands for the \texttt{Fanto10} set at $Q=1.6$ GeV, with $\alpha_s(M_Z)=0.118$.}
    \label{fig:FantoPI-nuclear}
\end{figure}

Among $\sim 100$ acceptable models, we find the best fit corresponding to the $\chi^2$ minimum, 
$\chi^2_0\equiv \min_{\{M\}} \chi^2_{D|M},$ 
and keep other solutions  satisfying
\begin{equation}
    \chi^2_{D|M}(M) \leq \chi^2_0 + \sqrt{2(N_{pts}-N_{par})},
    \label{eq:chi2LLM}
\end{equation}
where $\sqrt{2(N_{pts}-N_{par})}$ is one standard deviation on $\chi^2_{D|M}$ with $N_{pts}-N_{par}$ degrees of freedom. In the current analysis, we selected $\tilde{N}=5$ final models based on Eq.~(\ref{eq:chi2LLM}) and ad-hoc metric estimate to select the most diverse shapes. The $\tilde{N}$ models were then combined using the \METAPDF technique with the updated  \texttt{mcgen} \cite{Hou:2016sho} and \texttt{mp4lhc} packages~\cite{Gao:2013bia} adapted for the pion quark contents. In a nutshell, Hessian uncertainties were generated for each model according to the $\Delta \chi^2=1$ criterion \cite{Collins:2001es}, then converted into  MC replicas, which were then bundled, in an unweighted way, to form the combined MC ensemble. The latter, in turn, was converted into a Hessian ensemble called \FantoPDF, with 15 eigenvector sets (1 set per eigenvector direction). 
Figure~\ref{fig:FantoPI-nuclear} displays the resulting PDFs and uncertainties for three flavor combinations --- valence $xu_V=xV/2$, $x\bar u=xS/6$, and gluon $g$ --- as solid orange bands. 

As a resampling method, the \METAPDF combination ensures that the MC PDFs are identically generated for all models' outputs. The ${\tilde N}$ models are exchangeable (the combination does not depend on the order in which the models are added), while the \Fantomas methodology guarantees that all input Hessian sets are converted into error PDF bundles according to an identical procedure. %
PDF inference in the \Fantomas framework thus provides an ensemble of  i.i.d. MC sets, %
and as a result the \Fantomas methodology, followed by the combination via the \METAPDF method, renders an estimate of epistemic uncertainty that improves as more functional forms are explored. The presented implementation sufficient for pion fits, which nevertheless goes beyond the previous results utilizing fixed parametrization forms, can be further augmented with features already deployed in precision fits, such as closure tests \cite{Harland-Lang:2024kvt}, cross validation and expanded goodness-of-fit criteria \cite{Kovarik:2019xvh}. 

{\it Including nuclear uncertainty in pion PDFs.} The \Fantomas framework can handle other types of uncertainties, like the one coming from nuclear PDFs in the key Drell-Yan data sets from E615 and NA10. 
%
\iffalse
The leading-power collinear factorization for DY pair production cross section is
%
\begin{eqnarray}
\label{eq:DYfactorization}
%
\sigma &=&{} \sum_{a,b}\int\!dx_a \int\!dx_b f_{a/A}(x_a,\mu_\mathrm{F}^2)\\
&&
\, f_{b/B}(x_b,\mu_\mathrm{F}^2)
H_{a,b,x_a,x_b,\mu_\mathrm{F}^2}
+ {\cal O}\!\left({M}/{Q}\right)
\;, \nonumber
\end{eqnarray}
%
where $f_{a/A}(x)$ is the pion PDF for a parton $a$ in a target $A=\pi^{\pm}$ and $f_{b/B}(x)$ is the proton PDF for parton $b$ in nuclear target $B$.
\fi
%
Given their size of the data set and impact, studies of nuclear PDFs (nPDFs) are currently more phenomenologically mature and precise than those of pion PDFs~\cite{Klasen:2023uqj, Kovarik:2015cma, Eskola:2021nhw, AbdulKhalek:2022fyi}. 
As the impact of pion data on the employed nPDF ensemble, here taken to be \nCTEQ NLO \cite{Kovarik:2015cma}, would be at most a small perturbation, one can capture constraints of the pion data on nPDFs by PDF reweighting,  without resorting to a simultaneous fit of nuclear and pion data. The statistical implementation in the \nCTEQ study, such as their tolerance factor,  QCD parameter values, and inference procedure, must be reconciled with the innovative features of our study. When estimating the total (pion+nuclear) uncertainty in \texttt{Fanto10}, we treat the \nCTEQ uncertainty as its authors suggest, namely, that the \nCTEQ Hessian eigenvector sets characterize an approximately normal probability distribution with respect to nuclear PDF parameters in the vicinity of the best fit given by the central \nCTEQ set. 

\begin{figure}
    \centering
    \includegraphics[width=0.9\linewidth]{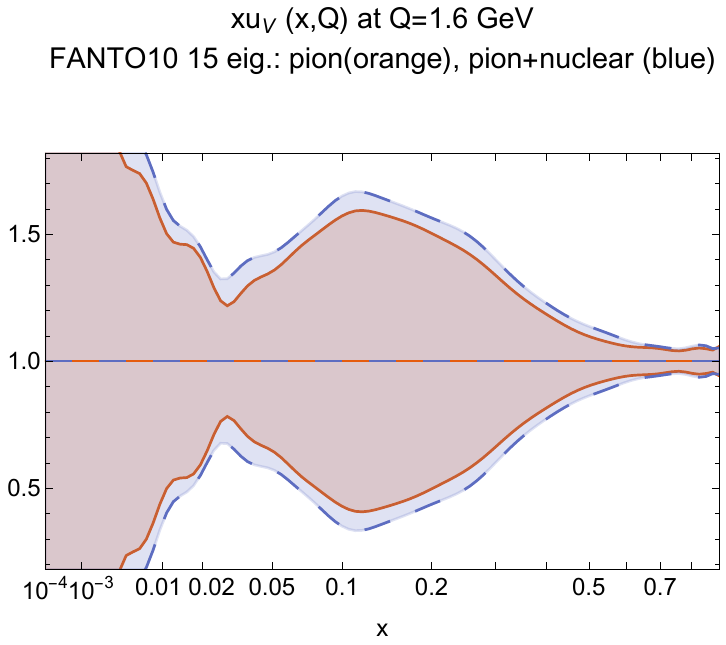}
    \caption{Aleatoric+epistemic (orange) and aleatoric+epistemic+nuclear (blue) uncertainty bands for the valence up quark PDF at $Q=1.6$ GeV, normalized to the respective central PDF.}
    \label{fig:FantoPI-nuclear_ratio}
\end{figure}

Our central assumption is that the pion parametrization and nuclear uncertainties are independent. Under this assumption, we combine the two uncertainties using the \METAPDF technique~\cite{Gao:2013bia}, as described above, with some additional steps:
\begin{enumerate}
\item The central lines and orange-band results shown in Fig.~\ref{fig:FantoPI-nuclear} correspond to using the Drell-Yan cross section with the central PDF set (replica 0, $f^{(0)}$) of the nPDF ensemble.  The central prediction is obtained by averaging the central predictions of $\tilde{N}$ PDF models, and the error bands are obtained as a bundle of $\tilde{N}$ respective error bands. %
\item Next, to estimate the nPDF uncertainty, we generate $N_{\rm nMC}=100$ nuclear MC (nMC) PDF replicas 
from $2D+1$ member sets of the \nCTEQ Hessian ensemble~\cite{Gao:2013bia,Hou:2016sho}.
A $k^{\rm th}$ replica, $f^{(k)}$, is computed by drawing a random vector $\vec R^{(k)}=\{R_1^{(k)},\, R_2^{(k)}, ..., R_D^{(k)}\}$ from a $D$-dimensional standard normal distribution:
\begin{eqnarray}
 f^{(k)} &=& f^{(0)} + d^{(k)} - \Delta, \mbox{ where}  \label{eq:fk}\\
    d^{(k)} &\equiv&  \sum_{i=1}^D \left( \frac{f^{(k)}_{+,i}-f^{(k)}_{-,i}}{2}  R_i^{(k)} \right.\nonumber\\
    &&\left.
    + \frac{f^{(k)}_{+,i}+f^{(k)}_{-,i}- 2 f^{(0)}}{2}  (R_i^{(k)})^2
    \right), \label{eq:dk}
\end{eqnarray}
and optional
$
    \Delta \equiv \langle d \rangle,
$ with $\langle \cdot \rangle$ the average over all replicas~\cite{Hou:2016sho}.
\item We repeat the pion \textit{central} fit for $N_{\rm nMC}$ nPDF replicas and $\tilde{N}$ pion parametrization models that were selected when fitting with \nCTEQ replica~0. From the resulting $N_{\rm nMC}\times \tilde{N}=500$ replicas, we keep only 396 that satisfy $\chi^2< \chi_0^2+\sqrt{2(N_{dat}-N_{par})}$, similarly to the criteria used by reweighting techniques~\cite{Ball:2010gb, Ball:2011gg}. These central fits quantify fluctuations in the nuclear PDFs. Since the fitting procedure, here in \xFitter, imposes physical constraints such as sum rules, we check that integrals of every generated PDF over the covered interval $x\in~[10^{-3}, 1]$ comply with the momentum sum rule at the level of about 1\%, and valence sum rule at the level better than 10\%.
\item Next, we add aleatoric and nuclear uncertainties in each of $\tilde{N}$ baseline pion models $M_i$. The former is taken to be the same as in the baseline fit with the central \nCTEQ replica (found using the $\Delta \chi^2=1$ prescription). The latter is computed from variations of central fits for $M_i$ when the \nCTEQ replica 0 is replaced by an alternative nMC replica generated in step 3. In practice, when generating the \METAPDF MC replicas from $M_i$, we replace Eq.~(\ref{eq:fk}) by
\begin{equation}
 f^{(k)} = f^{(0,k)}_{{\rm nMC}|M_i} + d^{(k)} - \Delta,  \label{eq:fkmod}\\
\end{equation}
with $f^{(0,k)}_{{\rm nMC}|M_i}$ being a randomly drawn nMC central replica (different for each $k$) from the ones generated for model $M_i$.
With the rest of MC generation staying the same, fluctuations of $f^{(k)}$ in $M_i$ are now enhanced by random contributions from the nMC uncertainty transmitted via $f^{(0,k)}_{{\rm nMC}|M_i}$.
\item The new MC sets for the ${\tilde N}$ models can be now combined and converted into a Hessian set with the \METAPDF algorithm, following~\cite{Gao:2013bia}. 
\end{enumerate}
The resulting Hessian PDFs, \nFantoPDF, now  include the nuclear uncertainty through the shifts of the input MC replicas in the previous steps. For the representative PDFs in Fig.~\ref{fig:FantoPI-nuclear}, the \nFantoPDF error bands are shown in blue color. The uncertainty coming from the convolution with nPDFs is moderate, but not entirely negligible compared to the intrinsic fitting uncertainty shown in orange, especially for the valence $xV(x,Q)$ that is better constrained. For $xV$, we can better visualize the uncertainties without and with the nuclear component by plotting them as ratios to the central PDF set in Fig.~\ref{fig:FantoPI-nuclear_ratio}. As the sea and gluon PDFs, $xS$ and $xg$, remain poorly known, their nuclear uncertainty can be neglected for the time being. 

%
%

\iffalse 
By default of the \xFitter package, \nCTEQ~\cite{Kovarik:2015cma} has been employed. Other choices could have been  the EPPS~\cite{Eskola:2021nhw} or the nNNPDF~\cite{AbdulKhalek:2022fyi} set. %
\fi
%

\begin{figure}
    \centering
    \includegraphics[width=0.8\linewidth]{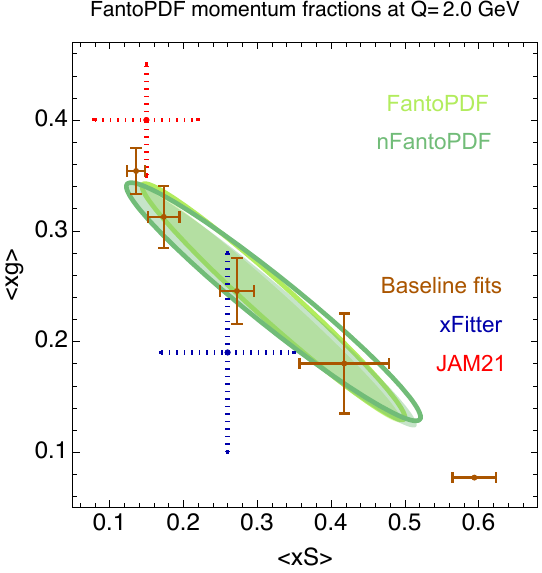}
    \includegraphics[width=0.9\linewidth]{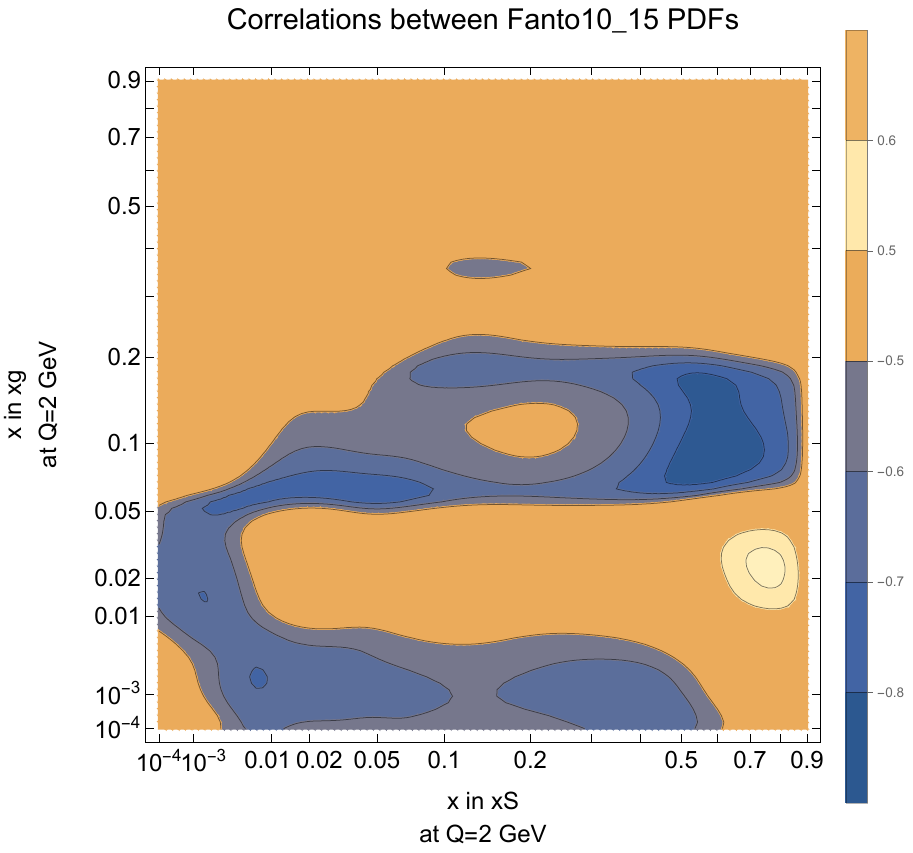}
    \caption{Top: Pion sea and gluon momentum fractions obtained with five \Fantomas baseline fits (brown error bars) and combined \FantoPDF and \nFantoPDF ensembles at $1\sigma$ level. We also show respective results from the JAM21 \cite{Barry:2021osv} (in red) and \xFitter (in blue) studies \cite{Novikov:2020snp}. See discussion in \cite{Kotz:2023pbu}. Bottom: The correlation cosine \cite{Nadolsky:2008zw} between $x S(x_S, Q)$ and $x G (x_g, Q)$ for \FantoPDF PDFs. We use $Q=2 \mbox{ GeV}$  in both panels.}
    \label{fig:FantoPI-nuclear_corr}
\end{figure}

{\it Phenomenology implications.} 
More complete estimation of the PDF parametrization dependence modifies correlation patterns among pion PDFs, both without and with the nuclear uncertainty. Among these, we confirm a strong anticorrelation between the gluon and sea PDFs and their momentum fractions, as depicted in the upper panel of Fig.~\ref{fig:FantoPI-nuclear_corr}. Since the dominant available data stem from $\pi N$ Drell-Yan pair production, which is primarily sensitive to a mix of quark and antiquark contributions, the gluon and sea-quark components remain entangled, with no clear sensitivity pattern currently allowing for their separation~\cite{KotzThesis:2025}. With \texttt{Fanto10\_(n)15} Hessian PDFs, we can trace how the anticorrelation originates from specific intervals of $x$. The lower panel of Fig.~\ref{fig:FantoPI-nuclear_corr} depicts a heat map of the correlation cosine \cite{Nadolsky:2008zw} between $xS(x_S,Q)$ and $xg(x_g,Q)$ in the plane of respective momentum fractions $x_S$ and $x_g$. While the gluon at $x_g>0.2$ is moderately correlated with $xS$ at all $x_S$, we see a high anticorrelation between $xg$ and $xS$ at $x_g<0.2$, producing the observed anticorrelation between momentum fractions $\langle xS\rangle$ and $\langle xg\rangle$ in the upper Fig.~\ref{fig:FantoPI-nuclear_corr}.

This degeneracy reflects the weakness of experimental constraints on the gluon and sea, with even a zero gluon contents still allowed by the data at the low scale of order $Q=1-2$ GeV. (See the detailed discussion in \cite{Kotz:2023pbu}.)
The degeneracy could be lifted with additional data at $x<0.1$ and from a broader range of processes --- see, {\it e.g.},\cite{Barry:2018ort, AbdulKhalek:2021gbh, Accardi:2023chb} --- thereby helping to answer the longstanding question of the pion's gluon content. 
COMPASS++'s range in $x$ will extend that on the old NA10 and E615 sets, possibly covering the mid-$x$ region~\cite{Andrieux:DIS24,Hsieh:HadPhys24}.
Looking forward, the Electron-Ion Collider (EIC) may provide valuable insights via  leading-neutron DIS~\cite{Arrington:2021biu}, although such measurements will require reliable theoretical input for the pion flux in the proton. Jefferson Lab, operating at 22 GeV, could also contribute to this effort, albeit with different nuclear uncertainties to be considered~\cite{Accardi:2023chb}.
Less conventional approaches include studies involving $J/\psi$ production, using historical data from experiments such as NA3 or E0705~\cite{Chang:2020rdy}.

{\it Conclusions.}
In this Letter, we have presented the final Hessian ensembles of NLO pion PDFs, \FantoPDF and \nFantoPDF, determined using the \Fantomas methodology. Our framework introduces a statistically motivated model combination to account for the uncertainty arising from the inference of the PDF parametrization~\cite{Courtoy:2024hog, Kotz:2023pbu}. We have further extended the analysis by incorporating an uncorrelated, third-party source of uncertainty propagated from the \nCTEQ ensemble of nuclear PDFs describing fitted cross sections from pion-tungsten Drell–Yan process.

The described methodology is broadly applicable to inverse problems involving multiple non-perturbative inputs. Examples include semi-inclusive deep inelastic scattering (SIDIS), where PDFs are convoluted with fragmentation functions, or leading-neutron DIS, where uncertainties in the pion flux must be taken into account. Our approach may also prove useful in the determination of nuclear PDFs themselves, where the baseline proton PDF typically lacks an associated uncertainty~\cite{Klasen:2023uqj}.
\section*{Acknowledgments}
We thank T. Hobbs, F. Olness, and CTEQ-TEA colleagues for various discussions.  
This study has been financially supported by 
a National Science Foundation AccelNet project, 
by the UNAM Grants No. DGAPA-PAPIIT IN111222 and IN102225,
by the U.S.\ Department of Energy under Grant No.~DE-SC0010129. 
PMN is grateful for support from Wu-Ki Tung Endowed Chair in particle physics.

\section*{Data Availability Statement}
Interpolation grids for the \FantoPDF and \nFantoPDF ensembles, without and with the nuclear uncertainty, respectively, can be downloaded from the CTEQ-TEA website \cite{CTPDFwebsite} and LHAPDF library \cite{LHAPDF6}. We use $\alpha_s(M_Z)=0.118$, and the grids cover the range of $1.3 \leq Q \leq 1000$ GeV and $0.0001\leq x\leq 1$, for the maximal active flavor number $N_f=5$. The supplemental material about \texttt{Fanto10} ensembles includes an extended collection \cite{CTPDFwebsite} of plots of PDF-PDF correlations (similar to the lower Fig.~\ref{fig:FantoPI-nuclear_corr}) and an examination of $L_2$ sensitivities \cite{Jing:2023isu} of fitted experiments to \FantoPDF PDFs in Chapter 4 of \cite{KotzThesis:2025}.
\bibliographystyle{utphys}
\bibliography{pionLett}

\end{document}